\documentclass[aps,prd,reprint,onecolumn,superscriptaddress,showpacs]{revtex4-1}
\usepackage{graphicx}
\usepackage{mathrsfs}
\usepackage{bm}
\usepackage{amsmath}
\usepackage{dcolumn}
\usepackage{epstopdf}
\usepackage{dsfont}
\usepackage{amssymb}
\usepackage{tabularx}
\usepackage{array}
\usepackage{float}
\usepackage{color}
\usepackage{epstopdf}
\usepackage{mathrsfs}
\usepackage[colorlinks, linkcolor=blue,anchorcolor=blue,citecolor=blue,urlcolor=blue]{hyperref}

\begin{document}
\title{Two-dimensional hourglass Weyl nodal loop in monolayer Pb(ClO$_{2}$)$_{2}$ and Sr(ClO$_{2}$)$_{2}$}
\author{Xin-Yue Kang}
\affiliation{School of Physics, Northwest University, Xi'an 710127, China}
\affiliation{Shaanxi Key Laboratory for Theoretical Physics Frontiers, Xi'an 710127, China}

\author{Chunmei Zhang}
\affiliation{School of Physics, Northwest University, Xi'an 710127, China}

\author{Mingxing Chen}
\email{mxchen@hunnu.edu.cn}
\affiliation{School of Physics and Electronics, Hunan Normal University, Key Laboratory for Matter Microstructure and Function of Hunan Province, Key Laboratory of Low-Dimensional Quantum Structures and Quantum Control of Ministry of Education, Changsha 410081, China}

\author{Si Li}
\email{sili@nwu.edu.cn}
\affiliation{School of Physics, Northwest University, Xi'an 710127, China}
\affiliation{Shaanxi Key Laboratory for Theoretical Physics Frontiers, Xi'an 710127, China}
	
\begin{abstract}
The hourglass fermions in solid-state materials have been attracting significant interest recently. 
However, realistic two-dimensional (2D) materials with hourglass-shaped band structures are still very scarce. Here, through the first-principles calculations, we identify the monolayer Pb(ClO$_{2}$)$_{2}$ and Sr(ClO$_{2}$)$_{2}$ materials as the new realistic materials platform to realize 2D hourglass Weyl nodal loop. We show that these monolayer materials possess an hourglass Weyl nodal loop circling around the $\Gamma$ point and Weyl nodal line on the Brillouin zone (BZ)
boundary in the absence of spin-orbit coupling (SOC). Through the symmetry analysis, we demonstrate that the hourglass Weyl nodal loop and Weyl nodal line are protected by the nonsymmorphic symmetries, and are robust under the biaxial strains. When we include the SOC,
a tiny gap will be opened in the hourglass nodal loop and nodal line, and the nodal line can be transformed into the spin-orbit Dirac points. Our results provide a new realistic material platform for studying the intriguing physics associated with the 2D hourglass Weyl nodal loop and spin-orbit Dirac points.

\end{abstract}
	
\maketitle
\section{Introduction}
Topological quasiparticles formed by the band crossing points have been attracting extensive research interest in current condensed matter physics.
Different from elementary particles that are constrained by the Poincar\'e symmetry in high energy physics
, topological quasiparticles in condensed matter physics are constrained by the crystal space group symmetries which is much smaller subgroups of the Poincar\'e
symmetry~\cite{Bradley1972}. Therefore, besides the Weyl, Dirac, and Majorana fermions~\cite{armitage2018weyl,elliott2015colloquium,fu2008superconducting,murakami2007phase,wan2011topological,young2012dirac,wang2012dirac,wang2013three,niu2019mixed,zhao2020direct,zou2021antiferromagnetic} with high-energy analogs, a variety of novel unconventional topological quasiparticles have also been discovered in condensed matter physics.
For instance, band crossing may form higher-dimensional manifolds, leading to the nodal line~\cite{burkov2011topological,Weng2015c,Chen2015,Mullen2015,Fang2015,Yu2015,Kim2015a,Li2016,Bian2016,Huang2016,Yu2017,Li2017,wang2020ferromagnetic} and nodal surface~\cite{zhong2016towards,liang2016node,bzduvsek2017robust,wu2018nodal}, the multiple degenerate fermions (three-, six-, and eight-fold degenerate points)~\cite{weng2016topological,zhu2016triple,bradlyn2016beyond}, the higher-order fermions with quadratic or cubic dispersions~\cite{fang2012multi,yu2019quadratic}, the hourglass fermions with hourglass dispersions~\cite{wang2016hourglass}, and beyond~\cite{chen2017topological,kruthoff2017topological,yu2022encyclopedia}.

Hourglass fermion is the topological quasiparticles with hourglass-shaped dispersion in its band structures and usually is protected by the nonsymmorphic symmetries. It was first predicted theoretically on the surface bands of KHg$X$ ($X=$ As, Sb, Bi)~\cite{wang2016hourglass} material, and soon after was observed in the experiment by angle-resolved
photoemission spectroscopy (ARPES)~\cite{ma2017experimental,liang2017observation}. After that, some counterparts with hourglass bulk bands have also been proposed, such as hourglass Dirac chains and hourglass nodal loops~\cite{wang2017hourglass,takahashi2017spinless,li2018nonsymmorphic,wang2019hourglass,wu2020exhaustive,singh2018topological,zeng2020n,fu2018hourglasslike,fan2018cat}. 

 Due to there are more three-dimensional(3D) space groups (230 for 3D) than two-dimensional (2D) space groups (80 for 2D), so far, most discovered hourglass fermion materials are 3D, and it is difficult to realize the hourglass fermion in 2D materials. Only a few materials, including 2D $\mathrm{Bi} / \mathrm{Cl}$-$\mathrm{SiC}(111)$~\cite{wang2019hourglass} and monolayer GaTeI family~\cite{wu2019hourglass} have
been reported to be the 2D hourglass fermion materials. However, at present, the number
of 2D hourglass fermion materials is still very rare.  Furthermore, the proposed materials
also have their own disadvantages. For instance, the 2D $\mathrm{Bi} / \mathrm{Cl}$-$\mathrm{SiC}(111)$ material need the adsorption of atoms, and it is difficult to implement experimentally. For the monolayer GaTeI family materials, its hourglass-shaped bands host in the conduction band, and it cannot be detected directly. 
Hence, it is very desired to search more 2D materials that can realize hourglass fermions.

In this work, through first-principles calculations and symmetry analysis, we show monolayer Pb(ClO$_{2}$)$_{2}$ and Sr(ClO$_{2}$)$_{2}$ are the 2D hourglass fermion materials. We demonstrate that these monolayer materials are dynamically stable. These materials possess a hourglass nodal loop circling around the $\Gamma$ point in the valance band near the Fermi level in the absence of spin-orbit coupling (SOC). Since the SOC is not included, the nodal loop is twofold degenerate. It is thence be termed as the hourglass Weyl nodal loop. In addition, there is also a nodal line on the Brillouin zone (BZ) boundary with relatively flat in energy. We
clarify that the hourglass Weyl nodal loop and nodal line are protected by the glide mirror. The hourglass Weyl nodal loop and nodal line are robust under the uniaxial or biaxial strains. When the SOC is included, the hourglass Weyl nodal loop and nodal line open a tiny gap, and the nodal line on the BZ boundary transforms into the spin-orbit Dirac points located at three time reversal invariant momenta (TRIM) points. Our work provides a good material
system for exploring the intriguing physics of 2D hourglass Weyl nodal loop and spin-orbit Dirac points.

\section{FIRST-PRINCIPLES METHODS}
The first-principles calculations were based on the density functional theory (DFT), employing the projector augmented wave method as implemented in the Vienna \emph{ab initio} simulation package~\cite{Kresse1994,Kresse1996,PAW}. The Perdew-Burke-Ernzerhof-type generalized gradient approximation~\cite{PBE} was used for the exchange-correlation functional. The energy cutoff was set to 600 eV for a plane wave basis. The energy and force convergence criteria were set to be $10^{-7}$ eV and $0.005$ eV/\AA, respectively. The BZ was sampled with a $\Gamma$-centered $k$ mesh of size $10\times 10\times 1$. A vacuum layer with a thickness of 20 \AA\ was adopted to avoid artificial interactions between periodic images. The phonon spectrum was calculated using the PHONOPY code through the DFPT approach with $2\times 2\times 1$ supercell~\cite{togo2015first}. The irreducible representations (IRs) of the bands were calculated by the irvsp code~\cite{gao2021irvsp}.
\section{Crystal structure and stability}\label{structure}
\begin{figure}[htbp]
	\includegraphics[width=10cm]{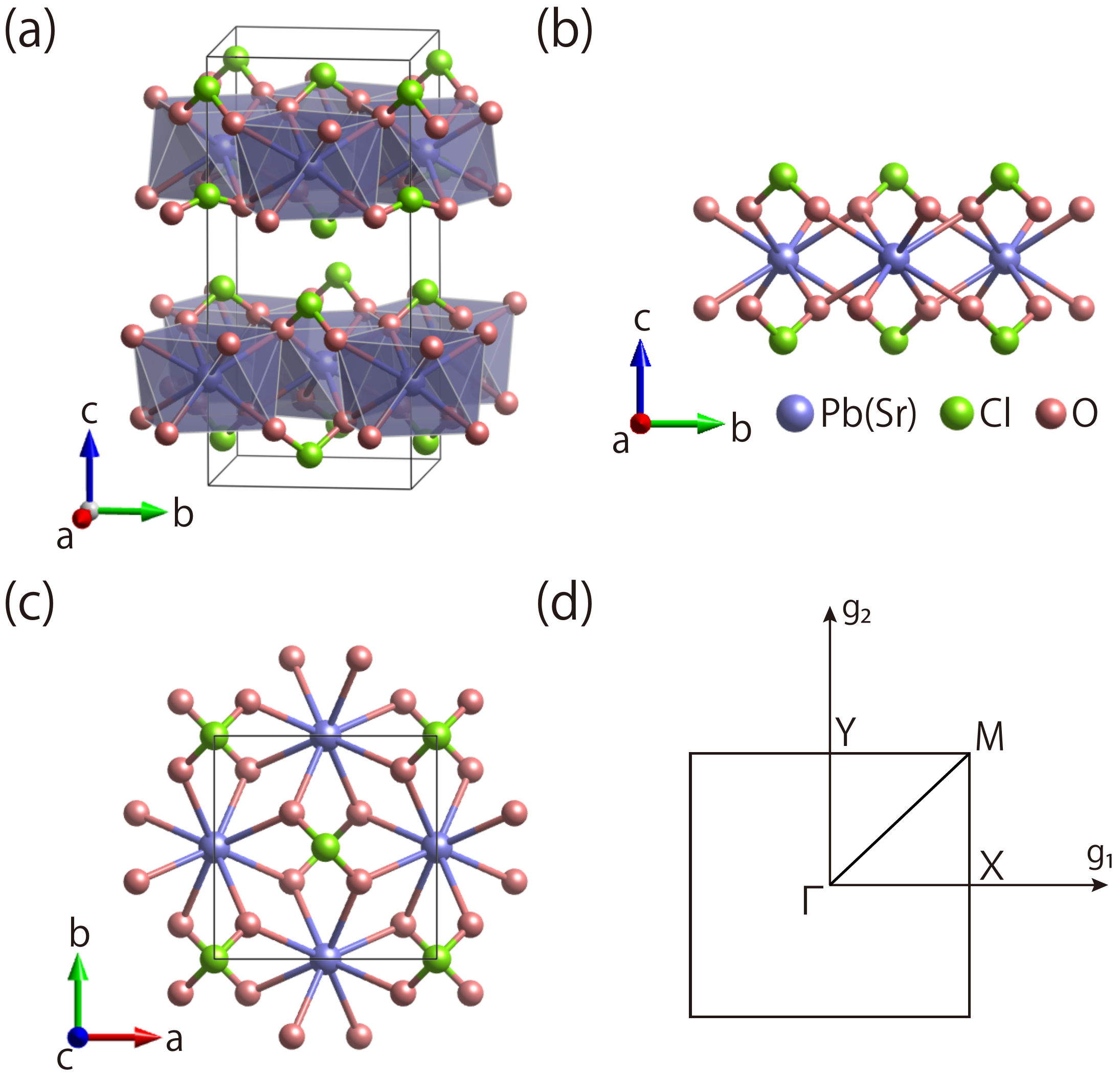}
	\caption{(a) Side view of the 3D bulk crystal structure of Pb(ClO$_{2}$)$_{2}$ and Sr(ClO$_{2}$)$_{2}$. The unit cell is shown by the black lines. (b) Side and (c) top view of the 2D monolayer structure
		of Pb(ClO$_{2}$)$_{2}$ and Sr(ClO$_{2}$)$_{2}$. The unit cell of the monolayer structure is indicated by the black lines in (c).
    (d) Brillouin zone of the monolayer structure. The high-symmetry points are given. 
		\label{fig1}}
\end{figure}

The 3D bulk of the Pb(ClO$_{2}$)$_{2}$ and Sr(ClO$_{2}$)$_{2}$ compounds were synthesized experimentally~\cite{okuda1990structures,smolentsev2005strontium}. The compounds crystallize in the orthorhombic space
group Ccca (No.68). The 3D bulk crystal structure is shown in Fig.~\ref{fig1}(a). It
is based on
separate layers parallel to the $ab$ plane, consisting of metal
cations that are coordinated by chlorite anions, and the O atoms
form almost ideal square antiprisms. Within the layers, each
anion bridges four metal cations~\cite{smolentsev2005strontium}. The experimental lattice
parameters for the Pb(ClO$_{2}$)$_{2}$ 3D bulk structure are $a=6.004$ \AA, $b=6.010$ \AA, and $c=12.504$ \AA; while for the Sr(ClO$_{2}$)$_{2}$ 3D bulk structure are $a=5.9799$ \AA, $b=5.9787$ \AA, and $c=12.7519$ \AA~\cite{okuda1990structures,smolentsev2005strontium}. Each unit cell in the 3D bulk crystal structure contains
two monolayer structures. The monolayer structures of Pb(ClO$_{2}$)$_{2}$ and Sr(ClO$_{2}$)$_{2}$ are shown in Figs.~\ref{fig1}(b) and (c), and their are obtained by cutting the 3D bulk along the direction of c axis. The monolayer Pb(ClO$_{2}$)$_{2}$ and Sr(ClO$_{2}$)$_{2}$ belong to the orthorhombic structure with space group Pban (No. 50) (it corresponding layer group of No. 39~\cite{po2017symmetry}). This space group includes the following three symmetry elements: the glide mirrors $\widetilde{\mathcal{M}}_{z}$:~$(x,y,z)\rightarrow(x+\frac{1}{2},y+\frac{1}{2},-z)$ and $\widetilde{\mathcal{M}}_{y}$:~$(x,y,z)\rightarrow(x+\frac{1}{2},-y+\frac{1}{2},z)$, the inversion $\mathcal{P}$ :~$(x,y,z)\rightarrow(-x+\frac{1}{2},-y+\frac{1}{2},-z)$. Here the tilde
denotes a nonsymmorphic operation, which involves a half lattice translation. By combining the 
three symmetry operations, we can obtain another glide mirror $\widetilde{\mathcal{M}}_{x}$:~$(x,y,z)\rightarrow(-x+\frac{1}{2},y+\frac{1}{2},z)$ (i.e. $\widetilde{\mathcal{M}}_{x}=\widetilde{\mathcal{M}}_{z}\widetilde{\mathcal{M}}_{y}\mathcal{P}$). In addition, since these material are 
non-magnetic, the time reversal symmetry $\mathcal{T}$ is also preserved. The optimized lattice
parameters for the monolayer structure Pb(ClO$_{2}$)$_{2}$ obtained from our
DFT calculations are $a=6.114$ \AA, $b=6.116$ \AA; while for the monolayer structure Sr(ClO$_{2}$)$_{2}$ are $a=6.046$ \AA, $b=6.045$ \AA. 

Since the band structures of Pb(ClO$_{2}$)$_{2}$ and Sr(ClO$_{2}$)$_{2}$ are very similar, we will concentrate on the results for Pb(ClO$_{2}$)$_{2}$ in the following discussion. The results for
Sr(ClO$_{2}$)$_{2}$ are shown in Appendix~\ref{appA}.

\begin{figure}[htbp]
	\includegraphics[width=10cm]{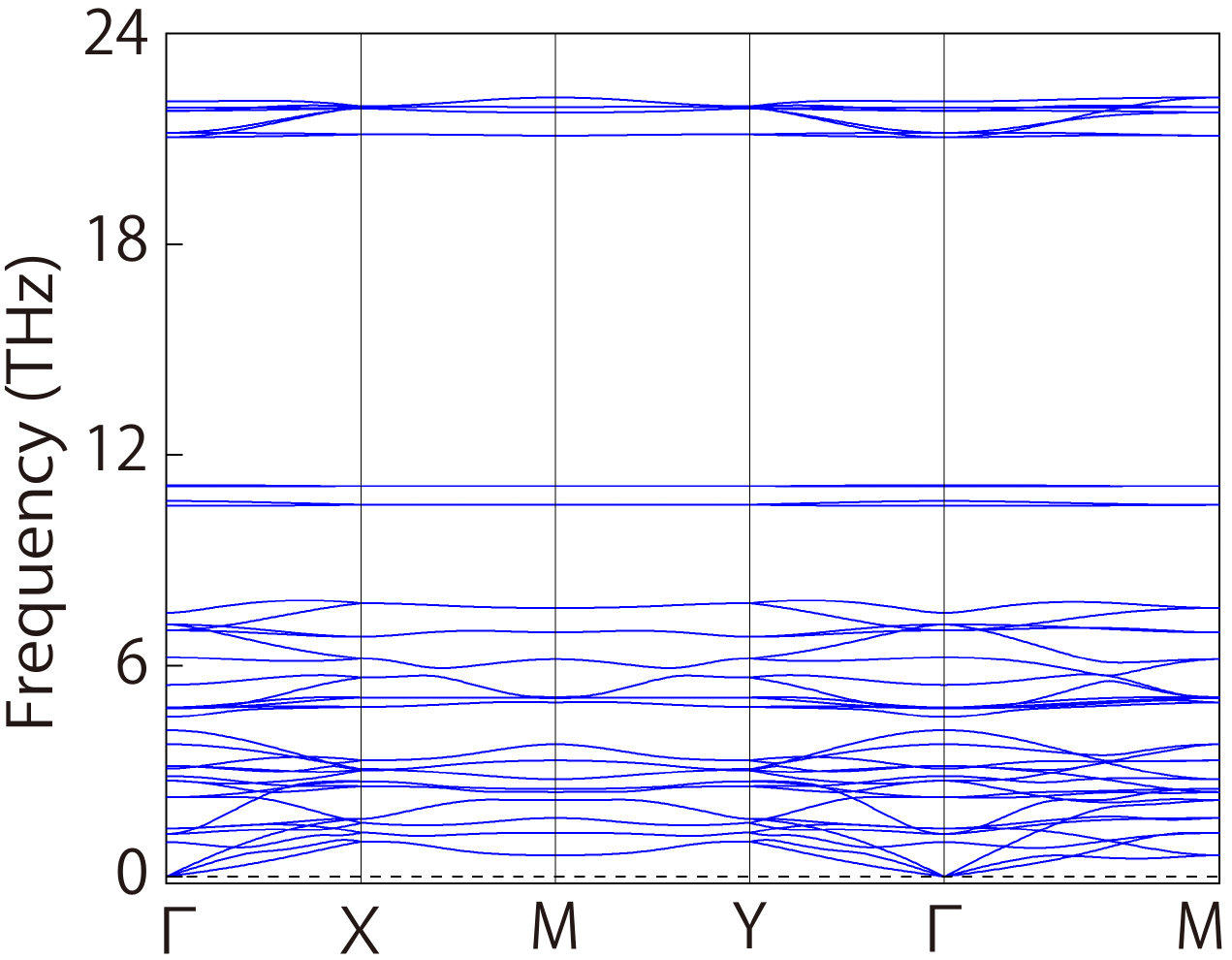}
     \caption{Phonon spectrum for the monolayer structure of Pb(ClO$_{2}$)$_{2}$.
	\label{fig2}}
\end{figure}

We first verify the stability of the monolayer structure of Pb(ClO$_{2}$)$_{2}$. To confirm its tability, we have calculated its phonon spectrum. The obtained phonon
spectrum is given in Fig.~\ref{fig2}, which shows that there is no
soft mode throughout the BZ, indicating that
the monolayer structure of Pb(ClO$_{2}$)$_{2}$ is dynamically stable. 

\begin{figure}[htbp]
	\includegraphics[width=10cm]{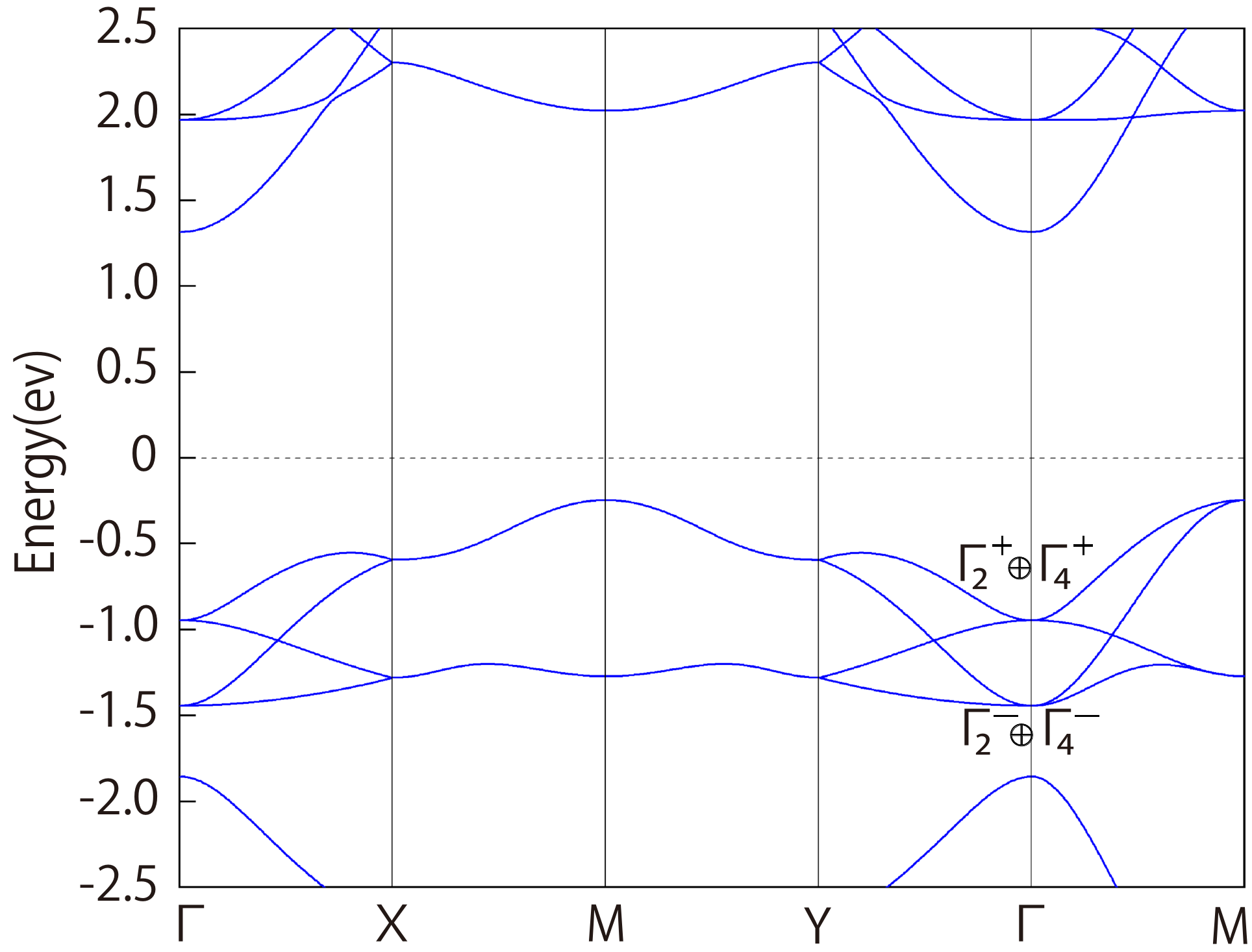}
	\caption{ Electronic band structure of monolayer Pb(ClO$_{2}$)$_{2}$ in the absence of SOC. The irreducible representations (IRs) of the valence bands near the Fermi level at $\Gamma$ point are given.
		\label{fig3}}
\end{figure}

\begin{figure}[htbp]
	\includegraphics[width=10cm]{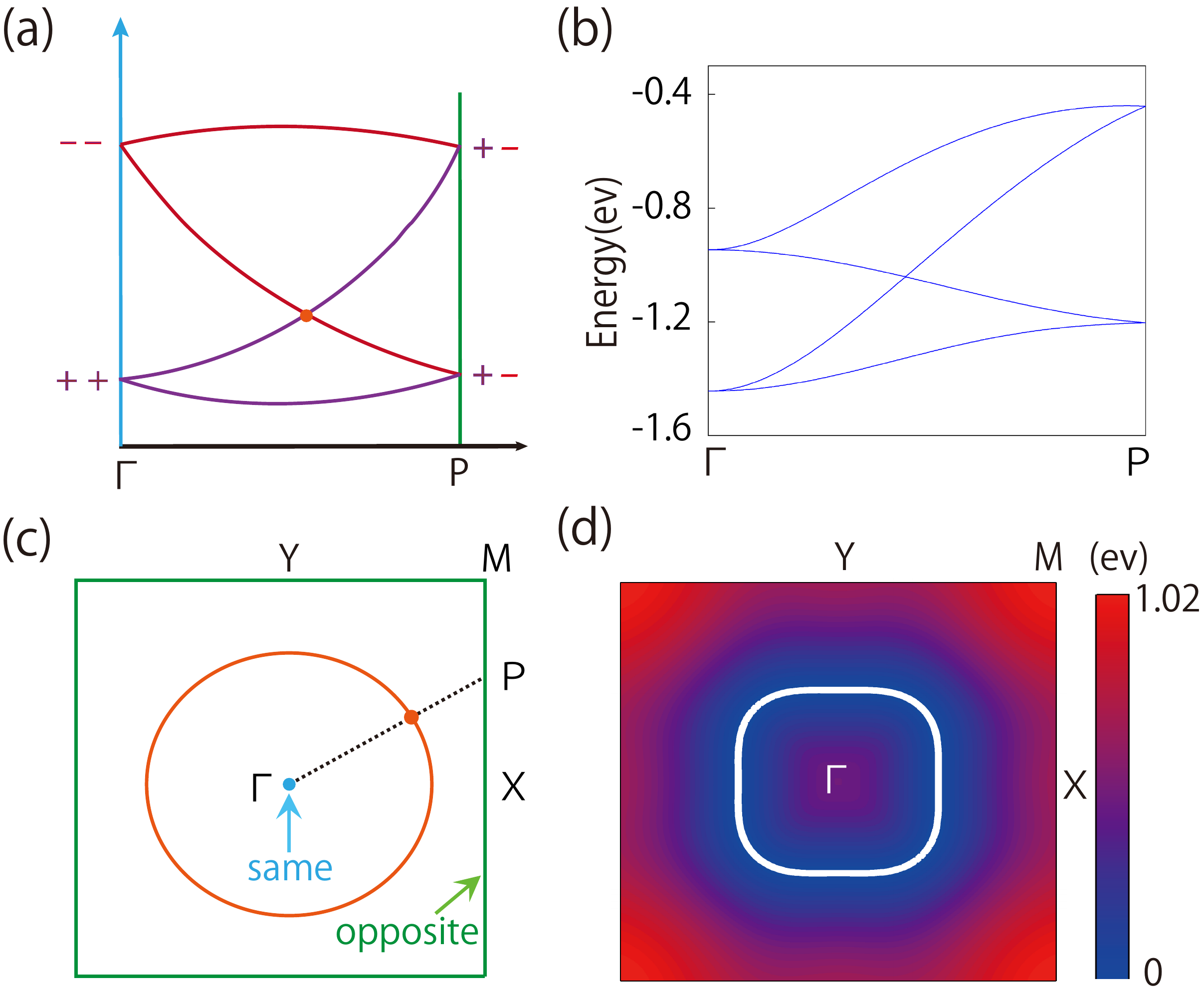}
	\caption{(a) Schematic diagram showing the hourglass-shaped dispersion
		along an arbitrary path connecting $\Gamma$ to the point
		P. (b) DFT calculation result for the
		hourglass dispersion along path $\Gamma$-P (P is the midpoint between
		X and M). (c) Schematic diagram showing that the neck point of
		the hourglass dispersion [the red point in (a)] traces out a twofold
		degenerate Weyl nodal loop around the $\Gamma$ point.
		(d) Shape of the hourglass nodal loop (the white colored
		loop) obtained from the DFT calculations. The color map indicates
		the local gap between the two crossing bands.
		\label{fig4}}
\end{figure}

\begin{figure}[htbp]
	\includegraphics[width=14cm]{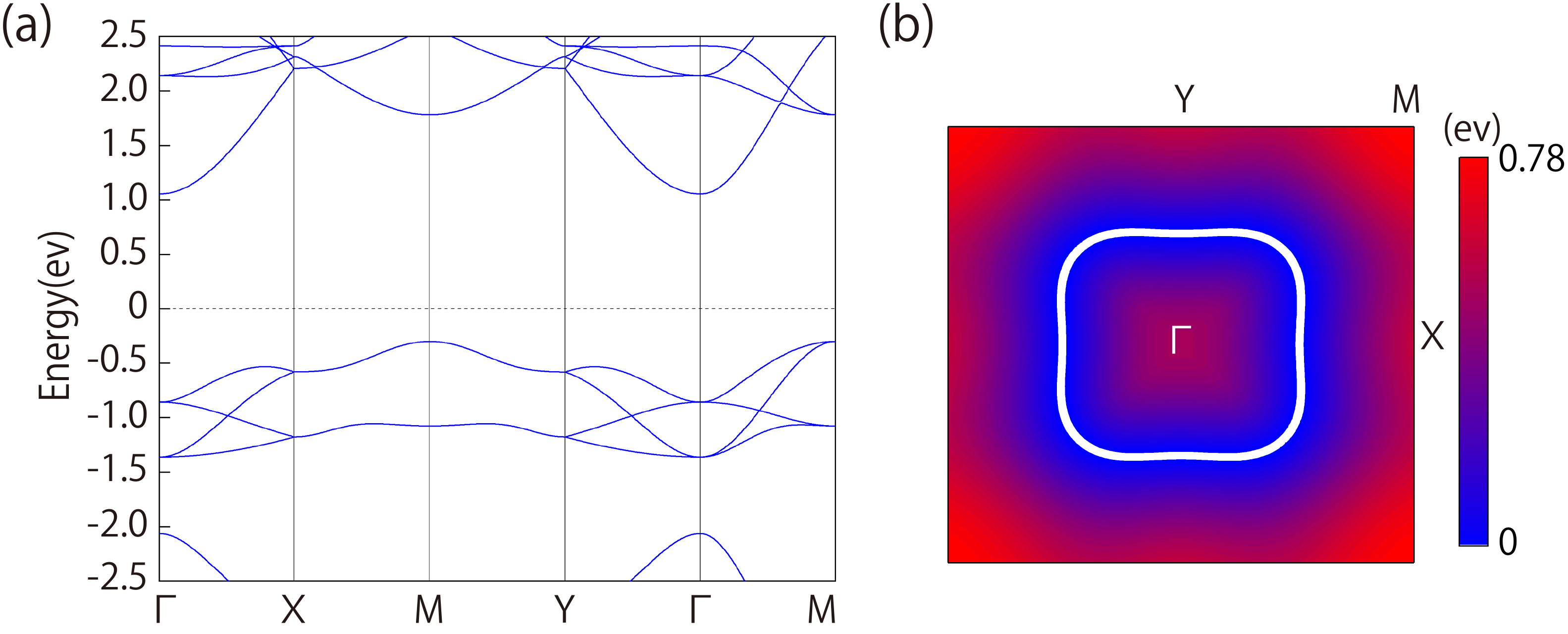}
	\caption{(a) Band structure of monolayer Pb(ClO$_{2}$)$_{2}$ in the absence of SOC  under +6\% biaxial strain.  Panel (b) shows
		the shape of the hourglass Weyl loop (white colored loop) under
		+6\% biaxial strain. The color map indicates
		the local gap between the two crossing bands.
		\label{fig5}}
\end{figure}

\section{Hourglass Weyl nodal loop and Symmetry analysis}
In this part, we consider the band structure of monolayer Pb(ClO$_{2}$)$_{2}$ without SOC, as shown in Fig.~\ref{fig3}. The band structure shows that the system is an indirect-gap semiconductor, and the band gap is about 1.56 eV. One can also observes that its valence band maximum (VBM) located at the M point, while its conduction band minimum (CBM) located at $\Gamma$ point. For the bands near VBM, there are several nontrivial features. First, there is an interesting hourglass-type dispersion on $\Gamma$-X, Y-$\Gamma$ and $\Gamma$-M. Actually, a careful scan indicates the neck crossing points of the hourglass dispersion on these paths are located on a nodal loop around the $\Gamma$ point, as shown in Fig.~\ref{fig4}(d). Second, we find that the
bands along the X-M and M-Y paths are twofold degenerate, which form a Weyl nodal line on the boundary of BZ. In addition, the dispersion
along the nodal line is very flat [see Fig.~\ref{fig3}].

Next, we will show that the hourglass Weyl nodal loop around the $\Gamma$ point and nodal line along the boundary of BZ are protected by the nonsymmorphic symmetries. 
 We noted that the 2D BZ is an invariant subspace of $\widetilde{\mathcal M}_z$, so each Bloch state $\left|u\right\rangle$ on the BZ can be chosen as an eigenstate of $\widetilde{\mathcal M}_z$. In addition, one can find that
\begin{equation}
	(\widetilde{\mathcal{M}}_{z})^2=T_{11}=e^{-i\left(k_{x}+k_{y}\right)}
\end{equation}
where $T_{11}$ is the translation along both $x$ and $y$ directions by a lattice constant. For the valance bands near the Fermi level at the $\Gamma$ point, the irreducible representations (IRs) of the bands calculated by our DFT are $\Gamma_{2}^{+}\oplus\Gamma_{4}^{+}$ and $\Gamma_{2}^{-}\oplus\Gamma_{4}^{-}$ [see Fig.~\ref{fig3}], which can allow the existence of twofold degeneracy.  The IRs $\Gamma_{2}^{+}\oplus\Gamma_{4}^{+}$ and $\Gamma_{2}^{-}\oplus\Gamma_{4}^{-}$ of $\widetilde{\mathcal M}_z$ respectively correspond to $\left(\begin{array}{ll} -1 & ~0 \\ ~0 & -1 \end{array}\right)$ and $\left(\begin{array}{ll} 1 & 0 \\ 0 & 1 \end{array}\right)$, which indicates that the twofold degenerate bands at $\Gamma$ point have the same $\widetilde{\mathcal M}_z$ eigenvalue $g_z=-1$ or $g_z=1$. 

Along the path X-M: $(\pi,k_y)$, the eigenvalues of $\widetilde{\mathcal M}_z$ are $g_z=\pm i e^{-ik_y/2}$. Note that each $k$ point on X-M is also invariant under the combined anti-unitary operation $\widetilde{\mathcal M}_y\mathcal T$. Since
	$(\widetilde{\mathcal M}_y\mathcal T)^2=e^{-ik_x}=-1$
on X-M, the bands have a Kramer-like double degeneracy, forming the nodal line along this path. In addition, the commutation relation between $\widetilde{\mathcal M}_z$ and $\widetilde{\mathcal M}_y$ is
\begin{equation}\label{MzMy}
	\widetilde{\mathcal M}_z\widetilde{\mathcal M}_y=T_{0 1}\widetilde{\mathcal M}_y\widetilde{\mathcal M}_z.
\end{equation}
On X-M, we have $\widetilde{\mathcal M}_z\widetilde{\mathcal M}_y=e^{-ik_y}\widetilde{\mathcal M}_y\widetilde{\mathcal M}_z$. Hence, for an eigenstate $|u\rangle$ with $\widetilde{\mathcal M}_z$ eigenvalue $g_z$, we have
\begin{equation}
	\widetilde{\mathcal M}_z(\widetilde{\mathcal M}_y\mathcal T)|u\rangle=-g_z(\widetilde{\mathcal M}_y\mathcal T)|u\rangle,
\end{equation}
indicating that the two Kramers partners $|u\rangle $ and $(\widetilde{\mathcal M}_y\mathcal T)|u\rangle$ have the opposite eigenvalue $g_z$. We can analyse in a similar way about the nodal line on the M-Y path, and the detailed
analysis is shown in Appendix~\ref{appB}. In Appendix~\ref{appB}, we shown that the nodal line on the M-Y path is protected by $\widetilde{\mathcal M}_x\mathcal T$ and two Kramers partners $|u\rangle $ and $(\widetilde{\mathcal M}_x\mathcal T)|u\rangle$ also have the opposite eigenvalue $g_z$. Therefore, all the BZ boundaries have a Kramer-like double degeneracy with the opposite eigenvalue $g_z$.  
Thus, there must be a partner-switching for the band structure when going from $\Gamma$ point to an arbitrary point (such as point P) on the BZ boundary, which forms the hourglass-type dispersion, as schematically illustrated in Fig.~\ref{fig4}(a). This argument is also confirmed by our DFT result as shown in Fig.~\ref{fig4}(b).

Consequently, as a result of the above analysis, the neck
point of the hourglass dispersion must trace out a Weyl nodal
loop centered around the $\Gamma$ point, as schematically shown in Fig.~\ref{fig4}(c).
This is also confirmed by our DFT calculation result [see Fig.~\ref{fig4}(b)]. Our analysis shows that this
hourglass Weyl nodal loop is protected by the
nonsymmorphic space group symmetries.

\section{Strain effect on band structure}
Two-dimensional materials have good mechanical properties, and strain engineering is a very powerful technique to tune the
properties of 2D materials. Due to the hourglass Weyl nodal loop around the $\Gamma$ point is protected by the $\widetilde{\mathcal M}_z$, $\widetilde{\mathcal M}_y\mathcal T$ and $\widetilde{\mathcal M}_x\mathcal T$ symmetries, it will cannot be destroyed when these symmetries are preserved. 
In fact, we noted that these symmetries can hold under some strains, such as in-plane biaxial and uniaxial strains. To 
verify this, the calculated
band structure of monolayer Pb(ClO$_{2}$)$_{2}$ under the +6$\%$ biaxial
strain is show in Fig.~\ref{fig5}(a). We indeed observes that the hourglass
Weyl nodal loop still exists [see Fig.~\ref{fig5}(a) and (b)].

\section{Effect of SOC}

\begin{figure}[htbp]
	\includegraphics[width=10cm]{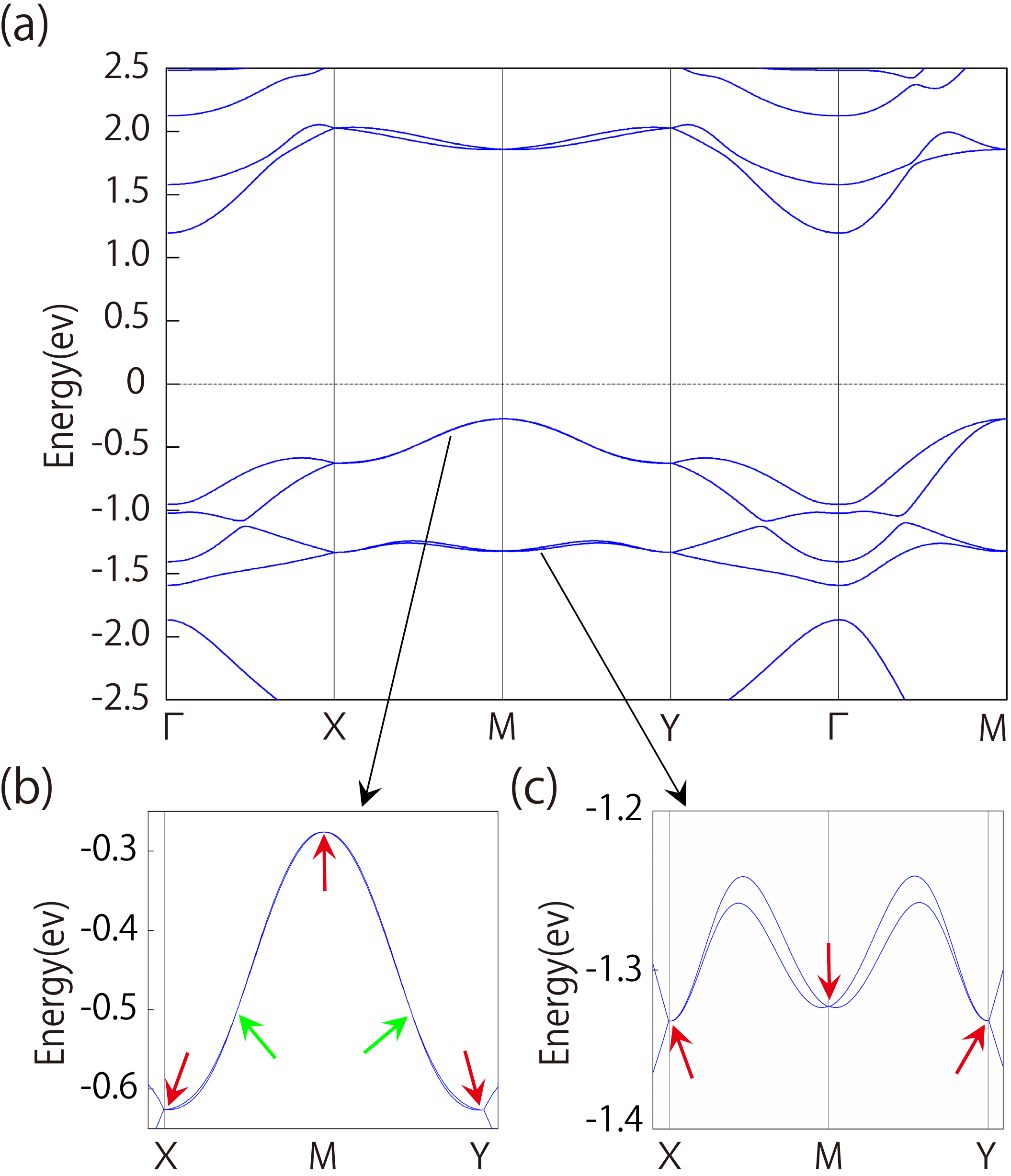}
	\caption{(a) Band structure of monolayer Pb(ClO$_{2}$)$_{2}$ with SOC included. (b) and (c) are the zoom-in band structures for the valance bands near the Fermi energy along X-M-Y. The red arrows show the Dirac points at X, M, and Y, and the green arrows in (b) show the two type-II Dirac points.
	\label{fig6}}
\end{figure}

The above discussion is not included SOC. Here, we
investigate its band structure in the presence of SOC, and the band
structure with SOC is show in Fig.~\ref{fig6}. It should be noted that the system has the inversion and time reversal symmetries. Hence, each band is at least twofold degenerate because of the presence of both time reversal and inversion symmetries [with $(\mathcal{PT})^2=-1$]. One observes that the monolayer Pb(ClO$_{2}$)$_{2}$ is
also an indirect-gap semiconductor and the band gap is about 1.47 eV. For
the bands near VBM, one can find that the hourglass Weyl nodal loop opens a small gap and the nodal lines along X-M and M-Y are also split. However, the band at TRIM points X, M, and Y points are fourfold degenerate, forming the Dirac points. The Dirac points at X, M, and Y are protected by the nonsymmorphic symmetries and are robust against SOC. Hence, they are spin-orbit Dirac points~\cite{young2015dirac,guan2017two}.
In the following, we show that the Dirac points at X and Y points are protected by the
following symmetries: glide mirror $\widetilde{\mathcal{M}}_{z}$, inversion $\mathcal{P}$, and time reversal $\mathcal{T}$. Each $k$-point in the 2D BZ is invariant under $\widetilde{\mathcal{M}}_{z}$, so each
Bloch state can be chosen as eigenstate of $\widetilde{\mathcal{M}}_{z}$. Because the SOC is included, we have
\begin{equation}
	(\widetilde{\mathcal{M}}_{z})^2=T_{11}\bar{E}=-e^{-i\left(k_{x}+k_{y}\right)}
\end{equation}
where $T_{11}$ denotes a lattice constant translation along both $x$ and $y$ directions
, and $\bar{E}$ is the $2\pi$ spin rotation. Hence the eigenvalues of $\widetilde{\mathcal{M}}_{z}$ are given by
\begin{equation}
	g_{z}=\pm i e^{-i k_{x} / 2-i k_{y} / 2}
\end{equation}
One notes that the commutation relation between $\widetilde{\mathcal{M}}_{z}$ and $\mathcal{P}$ is
given by
\begin{equation}
	\widetilde{\mathcal{M}}_{z} \mathcal{P}=T_{11} \mathcal{P} \widetilde{\mathcal{M}}_{z}=e^{-i k_{x}-i k_{y}} \mathcal{P} \widetilde{\mathcal{M}}_{z}
\end{equation}
As a result, for an eigenstate $|u\rangle$ with $\widetilde{\mathcal M}_z$ eigenvalue $g_z$, we have
\begin{equation}
	\widetilde{\mathcal M}_z(\mathcal{P}\mathcal{T})|u\rangle=-g_z(\mathcal{P}\mathcal{T})|u\rangle,
\end{equation}
This demonstrates that the two states $|u\rangle$ and $\mathcal{P}\mathcal{T}|u\rangle$
have opposite $\widetilde{\mathcal{M}}_{z}$ eigenvalues. One notes that X: $(\pi,0)$ and Y: $(0,\pi)$ are TRIM
points and $g_z= \pm 1$ is real at X and Y, hence any state $|u\rangle$ at X and Y have another Kramers partner $\mathcal{T}|u\rangle$ with the same eigenvalue $g_z$. This leads to four-fold
degeneracy at X and Y point, with the linearly independent states $\left\{|u\rangle, \mathcal{T}|u\rangle, \mathcal{P}\mathcal{T}|u\rangle, \mathcal{P}|u\rangle\right\}$.
The Dirac point at the M point can be analyzed in a similar way and the result is given in Appendix~\ref{appC}.

In addition, as indicated by the green arrow in Fig.~\ref{fig6}(b), we also find that there are two type-II Dirac points on the X-M and M-Y path, . Around the type-II Dirac point on the X-M path (M-Y path), the band is completely tipped over along the $k_y$-direction ($k_x$-direction) ~\cite{soluyanov2015type}. 

\section{Discussion and Conclusion}
In this work, we have revealed monolayer Pb(ClO$_{2}$)$_{2}$ and Sr(ClO$_{2}$)$_{2}$ as a good platform to study 2D hourglass Weyl nodal loop and 2D spin-orbit Dirac points. The hourglass Weyl nodal loop and spin-orbit Dirac points are determined by the nonsymmorphic symmetries of the system. Hence, the analysis in the main text can be applied for systems with similar symmetries, particularly for those 2D materials that belong to
space group No.50. In addition, due to the SOC splitting is very small for the band around the hourglass nodal loop in these monolayer materials [14 meV for Pb(ClO$_{2}$)$_{2}$ and 4 meV for Sr(ClO$_{2}$)$_{2}$], they
can be essentially regarded as 2D hourglass Weyl nodal loop materials. Since the hourglass Weyl nodal loop in the monolayer Pb(ClO$_{2}$)$_{2}$ and Sr(ClO$_{2}$)$_{2}$ is in the valence band near the Fermi level, it can be detected directly by ARPES. 

In conclusion, by first-principles calculation and theoretical analysis, we have proposed two realistic materials that realize 2D hourglass Weyl nodal loop and 2D spin-orbit Dirac points. These materials, monolayer Pb(ClO$_{2}$)$_{2}$ and Sr(ClO$_{2}$)$_{2}$, are shown to be stable in monolayer form, and host a hourglass Weyl nodal loop circling around the $\Gamma$ point in the valance band in the absence of SOC. In addition, there exists a nodal line on the BZ boundary. The hourglass Weyl nodal loop and nodal line are protected by the nonsymmorphic symmetries and are robust under various of applied strains. In the presence of SOC, the spin-orbit Dirac points appear at the three TRIM points. Our work provides a promising platform to explore the
fundamental physics of 2D hourglass Weyl nodal loop and 2D spin-orbit Dirac fermions.

\begin{acknowledgements}
This work is supported by the National Natural Science Foundation of China (Grants No. 12204378, No. 12004306, No. 12274342, No. 12174098, No. 11774084 and No. U19A2090) and project supported by State Key Laboratory of Powder
Metallurgy, Central South University, Changsha, China.
\end{acknowledgements}

\begin{appendix}

\section{Results for monolayer Sr(ClO$_{2}$)$_{2}$}\label{appA}
\begin{figure*}[htbp]
 	\includegraphics[width=18.6cm]{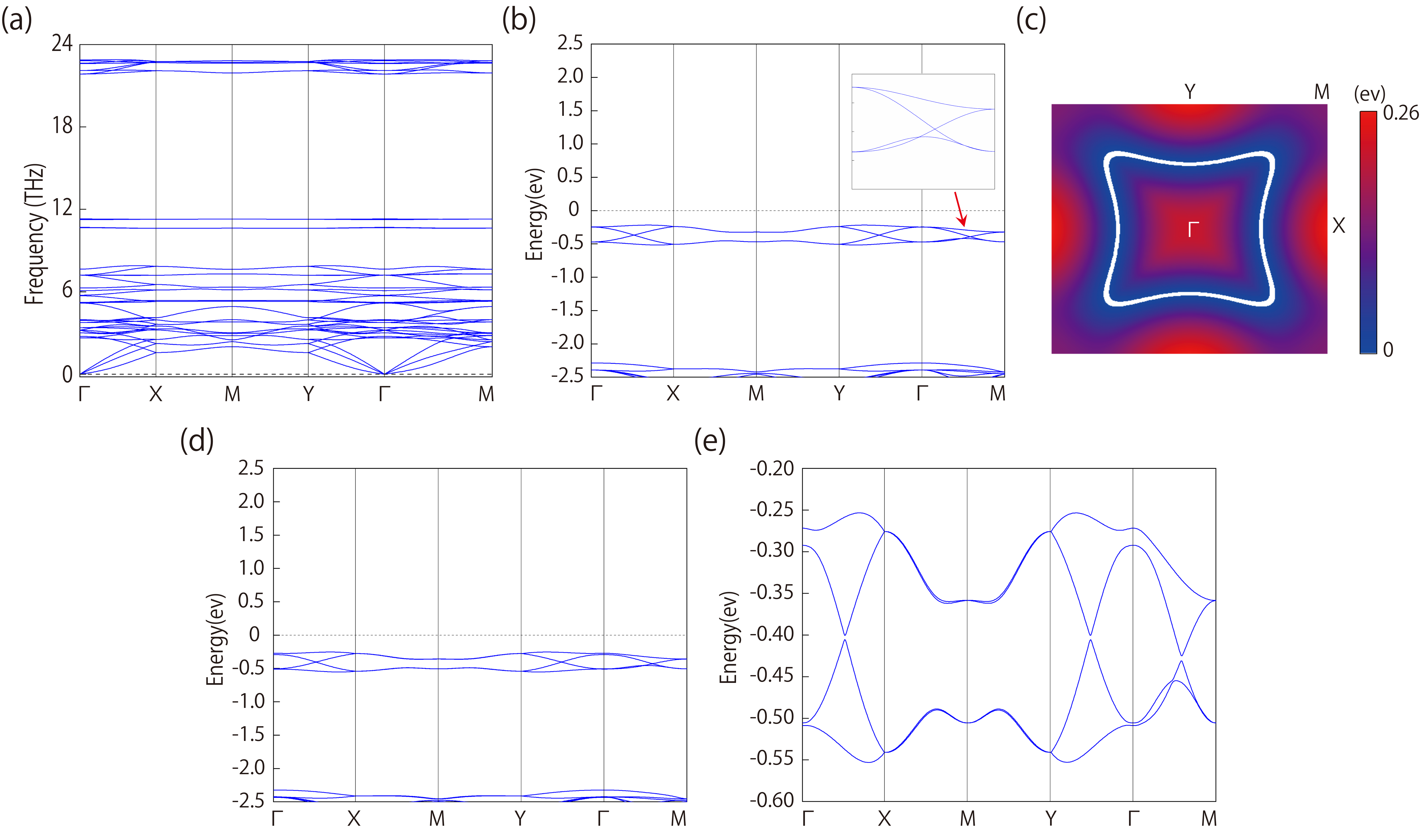}
 	\caption{(a) Calculated phonon spectrum for monolayer Sr(ClO$_{2}$)$_{2}$. (b)  Band structure of monolayer Sr(ClO$_{2}$)$_{2}$ in the absence of SOC. The zoom-in band structures for the low-energy bands
 		along $\Gamma$-M is given. (c) Shape of the hourglass nodal loop (the white colored
 		loop) obtained from the DFT calculations. The color map indicates
 		the local gap between the two crossing bands. (d) Band structure of monolayer Sr(ClO$_{2}$)$_{2}$ with SOC included. (e) The zoom-in band structures for the valance bands near the Fermi level.
 		\label{fig7}}
\end{figure*}

 Monolayer Sr(ClO$_{2}$)$_{2}$ has the same type of lattice structure as monolayer Pb(ClO$_{2}$)$_{2}$. From the calculated phonon
 spectrum [see Fig.~\ref{fig7}(a)], one can also find that the Sr(ClO$_{2}$)$_{2}$ monolayer structure is dynamically stable. The band structure without SOC for the monolayer structure is shown in Fig.~\ref{fig7}(b). One can observe that they are also possess hourglass-type dispersion for the valance bands near Fermi level, and the neck point of the hourglass dispersion form a Weyl nodal loop surrounding the $\Gamma$ point [see Fig.~\ref{fig7}(c)]. A degenerate and nearly flat band appears on the BZ boundary. When the SOC is included, the hourglass Weyl nodal loop around $\Gamma$ point and nodal line on the BZ boundary are split and three spin-orbit Dirac points appear at the three TRIM points X, M and Y [see Figs.~\ref{fig7}(d) and (e)]. Therefore, monolayer Sr(ClO$_{2}$)$_{2}$ share similar band structure features as those discussed in the main text for monolayer Pb(ClO$_{2}$)$_{2}$.
 
\section{Twofold degeneracy on M-Y path}\label{appB} 
Here, we show the detailed symmetry analysis of the degeneracy on M-Y when the SOC is not considered. Along the path M-Y: $(k_x,\pi)$, the eigenvalues of $\widetilde{\mathcal M}_z$ are $g_z=\pm i e^{-ik_x/2}$. Each $k$ point on M-Y is invariant under $\widetilde{\mathcal M}_x\mathcal T$. Since
$(\widetilde{\mathcal M}_x\mathcal T)^2=e^{-ik_y}=-1$
on M-Y, the bands along this path have a Kramer-like double degeneracy, leading to the nodal line along this path. Note that the commutation relation between $\widetilde{\mathcal M}_z$ and $\widetilde{\mathcal M}_x$ is
\begin{equation}\label{MzMx}
	\widetilde{\mathcal M}_z\widetilde{\mathcal M}_x=T_{1 0}\widetilde{\mathcal M}_x\widetilde{\mathcal M}_z.
\end{equation}
On M-Y, we have $\widetilde{\mathcal M}_z\widetilde{\mathcal M}_x=e^{-ik_x}\widetilde{\mathcal M}_x\widetilde{\mathcal M}_z$. Hence, for an eigenstate $|u\rangle$ with $\widetilde{\mathcal M}_z$ eigenvalue $g_z$, we have
\begin{equation}
	\widetilde{\mathcal M}_z(\widetilde{\mathcal M}_x\mathcal T)|u\rangle=-g_z(\widetilde{\mathcal M}_x\mathcal T)|u\rangle,
\end{equation}
indicating that the two Kramers partners $|u\rangle $ and $(\widetilde{\mathcal M}_x\mathcal T)|u\rangle$ have the opposite eigenvalue $g_z$.
 
\section{Dirac point at the M point}\label{appC}
The point M: $(\pi,\pi)$ is invariant under the symmetries $\widetilde{\mathcal{M}}_{z}$, $\widetilde{\mathcal{M}}_{y}$, and $\mathcal{T}$. It should be noted that $(\widetilde{\mathcal{M}}_{y}\mathcal{T})^2=-1$ is still preserved when the SOC included.   This leads to a Kramers-like twofold degeneracy at M point. It is also important to note that for a state $|u\rangle$ at M with $\widetilde{\mathcal{M}}_{z}$ eigenvalue $g_z$, its Kramers partner $\mathcal{T}|u\rangle$ has the opposite eigenvalue $-g_z$ (due to $g_z=\pm i$ at M). In addition, with the SOC included, the commutation relation between $\widetilde{\mathcal M}_z$ and $\widetilde{\mathcal M}_y$ is
\begin{equation}\label{MzMy}
	\widetilde{\mathcal M}_z\widetilde{\mathcal M}_y=-T_{0 1}\widetilde{\mathcal M}_y\widetilde{\mathcal M}_z.
\end{equation}
where $T_{01}$ denotes a lattice constant translation along $y$ directions, and the negative sign is because the anticommutativity between
two spin rotations, i.e., from $\left\{\sigma_y,\sigma_z\right\}=0$.
Hence, at M point, one finds $\widetilde{\mathcal M}_z\widetilde{\mathcal M}_y=-\widetilde{\mathcal M}_y\widetilde{\mathcal M}_z$. Therefore, for an eigenstate $|u\rangle$ with $\widetilde{\mathcal M}_z$ eigenvalue $g_z$, we have
\begin{equation}
	\widetilde{\mathcal M}_z(\widetilde{\mathcal M}_y\mathcal T)|u\rangle=g_z(\widetilde{\mathcal M}_y\mathcal T)|u\rangle,
\end{equation}
which indicates that the two states $|u\rangle$ and $\widetilde{\mathcal{M}}_{y}\mathcal{T}|u\rangle$ have the same $g_z$.
As a result, the four states $\{|u \rangle, \mathcal{T}|u\rangle, \widetilde{\mathcal{M}}_{y}|u \rangle, \widetilde{\mathcal{M}}_{y}\mathcal{T}|u\rangle\}$ are linearly independent and degenerate with the same energy, forming the Dirac point.

\end{appendix}

%\bibliography{2DHourglass_NL__ref}


\end{document}